\documentclass[11pt,twoside]{article}
\usepackage{asp2010}

\resetcounters

\begin{document}

\title{Mass Loss and Variability in Evolved Stars}
\author{Mssimo~Marengo
\affil{Dept. of Physics and Astronomy, Iowa State University,
  Ames, IA (USA)}} 

\begin{abstract}
  Mass loss and variability are two linked, fundamental properties of
  evolved stars. In this paper I review our current understanding of
  these processes, with a particular focus on how observations and
  models are used to constrain reliable mass loss prescriptions for
  stellar evolution and population synthesis models.
\end{abstract}

\section{Introduction}
\label{sec:intro}

Stars are the alchemist's dream: they are the nuclear furnaces
transmuting the light nuclei from the dawn of the cosmos into the
heavier elements found in planets and living organisms. It is only
thanks to the process of \emph{mass loss}, however, that these
enriched elements are finally released to the InterStellar Medium
(ISM), where they can form a new generation of stars, closing the loop
in the galactic ecosystem. It is trough mass loss that stars became
the engines of galactic chemical evolution \citep[see
e.g.][]{tinsley1968, chiosi1986}.

While mass loss processes are active in most evolutionary phases, they
became especially efficient after the end of the main sequence. The
basic principles driving the intense mass loss of evolved stars have
been well recognized for over 4 decades \citep{salpeter1974, kwok1975,
  goldreich1976}: as stars switch from core to shell nuclear burning,
the added luminosity swells them to giant or supergiant radius,
leading to low gravity and cool effective temperature. Their
atmospheres are further destabilized by radial pulsations, induced as
the stars cross the Long Period Variability (LPVs) instability strip.
Shocks from the pulsations compress and levitate the tenuously bound
external atmospheric layers, triggering an outflow. Particulates
(astronomical dust) condensate \citep{sedlmayr1994} and are pushed by
stellar radiation pressure, dragging the molecular gas and further
enhancing the wind. Through this mechanism, mass loss rates can reach
values as high as $\sim 10^{-4} \ M_\odot$/yr \citep[see e.g. reviews
by][]{habing1996, willson2000}. As shown in
Figure~\ref{fig:HR-diagram}, this combination of low $g$, low
$T_{eff}$ and pulsations happens twice in the life of low and
intermediate mass stars ($M \la 10 \ M_\odot$), as they climb back the
Hayashi tracks during the Red Giant Branch (RGB) and the Asymptotic
Giant Branch (AGB) phases. High mass stars instead cross the LPV
instability strip as Red Supergiants (RSG).

\articlefigure[width=0.9\textwidth,angle=0]{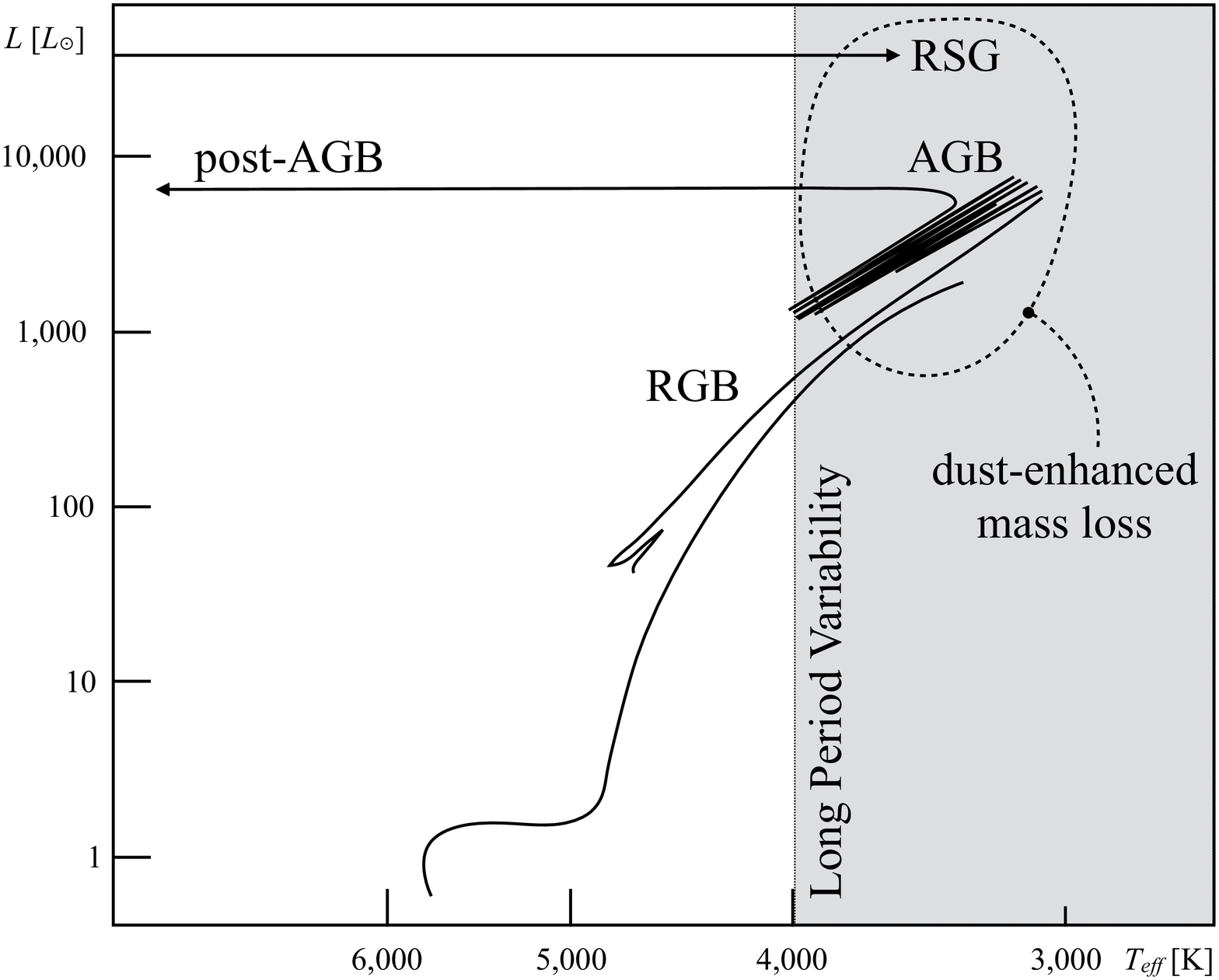}{fig:HR-diagram}
{Schematic view of a $1 \ M_\odot$ star late evolutionary phases
  characterized by intense mass loss and variability. Evolutionary
  track adapted from \citet{sackmann1993}. The trajectory of a RSG
  star is indicated at the top of the diagram.}

Despite the importance of mass loss for both stellar and galactic
evolution, many fine details of this process are unknown: we are
still lacking a comprehensive framework capable of predicting mass
loss rates from fundamental stellar parameters ($M$, $L$, $R$, [Fe/H]
and $\ell/H$). This lack of knowledge poses severe limitations to our
understanding of the last phases of stellar evolution: while the AGB
phase is driven by the mass of the inert C/O degenerate core
\citep{paczynski1970}, the end point is determined by the efficiency
of mass loss in depleting the convective envelope. Even a basic
parameter such as the mass above which a star ends its life as a
core-collapse supernova (SN), rather than an AGB, is very uncertain.
The long held assumption that all AGBs are progenitors of White Dwarfs
(WD) has recently been cast into doubt, with the hypothesis that
\emph{super}-AGBs (AGB stars in the 5-10 $M_\odot$ range) could be the
progenitors of dust-enshrouded, electron-capture SN
\citep{thompson2009, pumo2009}. Similarly, the mass range of intrinsic
carbon stars, and its dependence on zero age metallicity, has been the
subject of intense investigation. Despite a general agreement between
observations \citep[see e.g.][]{vanloon2005, feast2006,
  lebzelter2007}, and detailed evolutionary tracks \citep[see
e.g.][]{marigo2007}, both lower and upper mass limits for the
formation of carbon stars are still subject to revisions. At the lower
end, raising the C/O ratio above unity depends on the efficiency of
the third dredge-up during the Thermal Pulsing AGB phase (TP-AGB). At
the upper end the C/O ratio is reverted below unity by the Hot Bottom
Burning process \citep[HBB;][]{bloecker1991, boothroyd1992},
destroying the newly generated carbon in intermediate mass AGB stars.
Both limits are affected by the depletion of the stellar convective
envelope due to mass loss, which is uncertain. Finally, the
initial-final mass relation derived from the analysis of cluster WDs,
potentially providing strong constraints to the overall mass lost
during the RGB and AGB phases, still suffers from a large scatter
\citep[compare e.g.][]{catalan2008,kalirai2008, salaris2009}. All
these uncertainties pose serious difficulties for stellar population
synthesis simulations requiring accurate estimates of stellar yields
for galactic chemical evolution models.

In the last few years, however, significant progress has been made on
several fronts, from the availability of new instrumentation ideally
suited to measure the fine details of evolved star outflows, to the
coming of age of a new generation of dynamical models simulating the
processes at the base of mass loss, including the hydrodynamics of
pulsations, and the condensation of dust. In this paper I will discuss
the current state of affairs, with particular emphasis to the ongoing
quest for a parametric representation of mass loss that can be
employed in stellar evolution and population synthesis models.

\section{Observational Constraints}
\label{sec:obs}

Due to the large column density of dust enhanced winds, observation of
mass-losing evolved stars have traditionally been the domain of
infrared and sub-mm/radio telescopes \citep[see e.g.][for a
review]{marengo2009}. Radio observations, in particular, are sensitive
to the gas component of the outflow, and can provide reliable
estimates of the circumstellar \emph{gas} mass by spectrally resolving
rotational transitions of molecular tracers, such as CO \citep[see
e.g.][for a recent survey]{debeck2010}. These observations have the
advantage of directly measuring the outflow velocity, which is needed
to convert the mass of the circumstellar molecular envelope into a
mass loss rate (e.g. $M_{CO}$ into $\dot M_{CO}$), but still rely on
uncertain estimates of the molecular tracer abundance to derive the
total mass loss $\dot M$ of the star. Fitting the mid-IR excess with
radiative transfer models, on the other hand, provides robust
measurements of the circumstellar \emph{dust} mass \citep[see
e.g.][]{groenewegen2006, sargent2010, srinivasan2010}, but not of the
total mass loss rate, which still requires an independent knowledge of
the kinematics of the envelope (the wind velocity) and of the gas to
dust mass ratio. These free parameters are responsible for
uncertainties in the total mass loss rate as large as one order of
magnitude. Near-simultaneous radio/sub-mm/infrared observations
provide the most stringent constrains to the mass loss parameters, but
still suffer from the tendency of different tracers (multiple dust
components and different molecular transitions) to be spatially
segregated, and may probe different mass loss timescales.

\articlefigure[width=0.9\textwidth,angle=0]{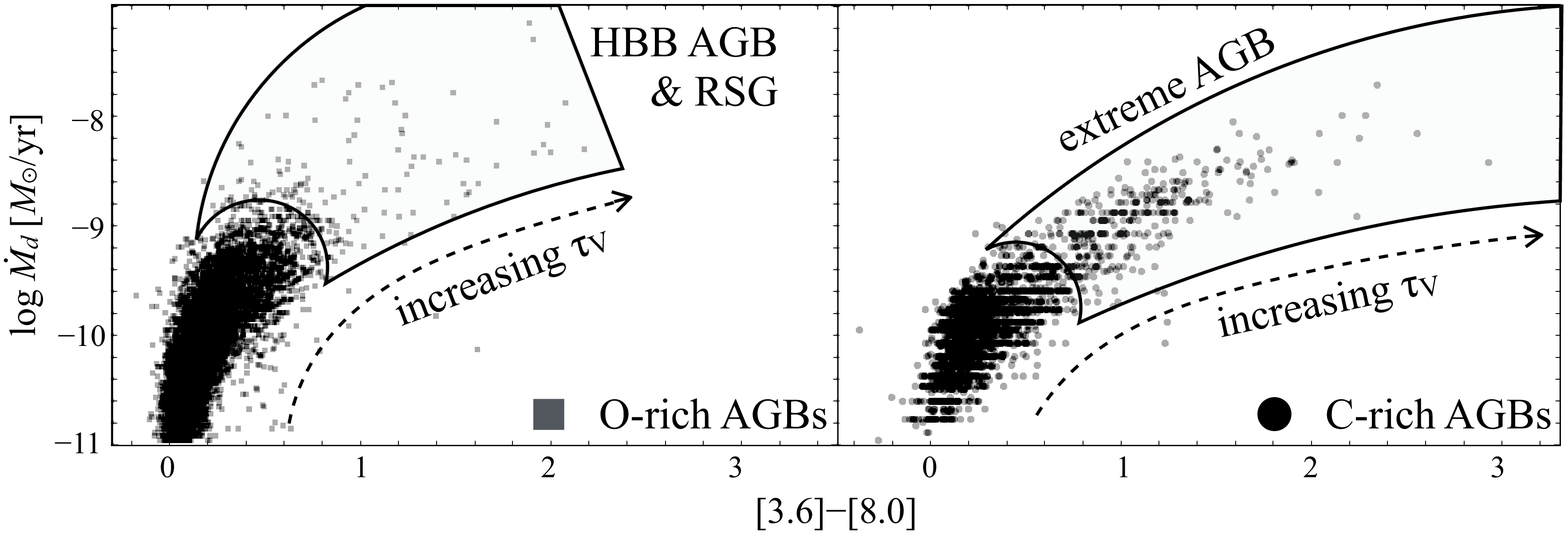}{fig:dmdt}
{Dust mass loss rate vs. [3.6]$-$[8.0] color for O-rich (\emph{left}) and 
  C-rich (\emph{right}) AGB stars in the LMC, as calculated by 
  \citet{riebel2012} by fitting SAGE photometry on MACHO selected 
  LPVs. Grey area indicate sources with high mass loss rate:
  intermediate mass AGBs undergoing HBB among the O-rich sources, and 
  the so-called ``extreme AGBs'' among the C-rich stars. RSGs would be 
  located in the same general area as the O-rich HBB sources.}

\articlefigure[width=0.90\textwidth,angle=0]{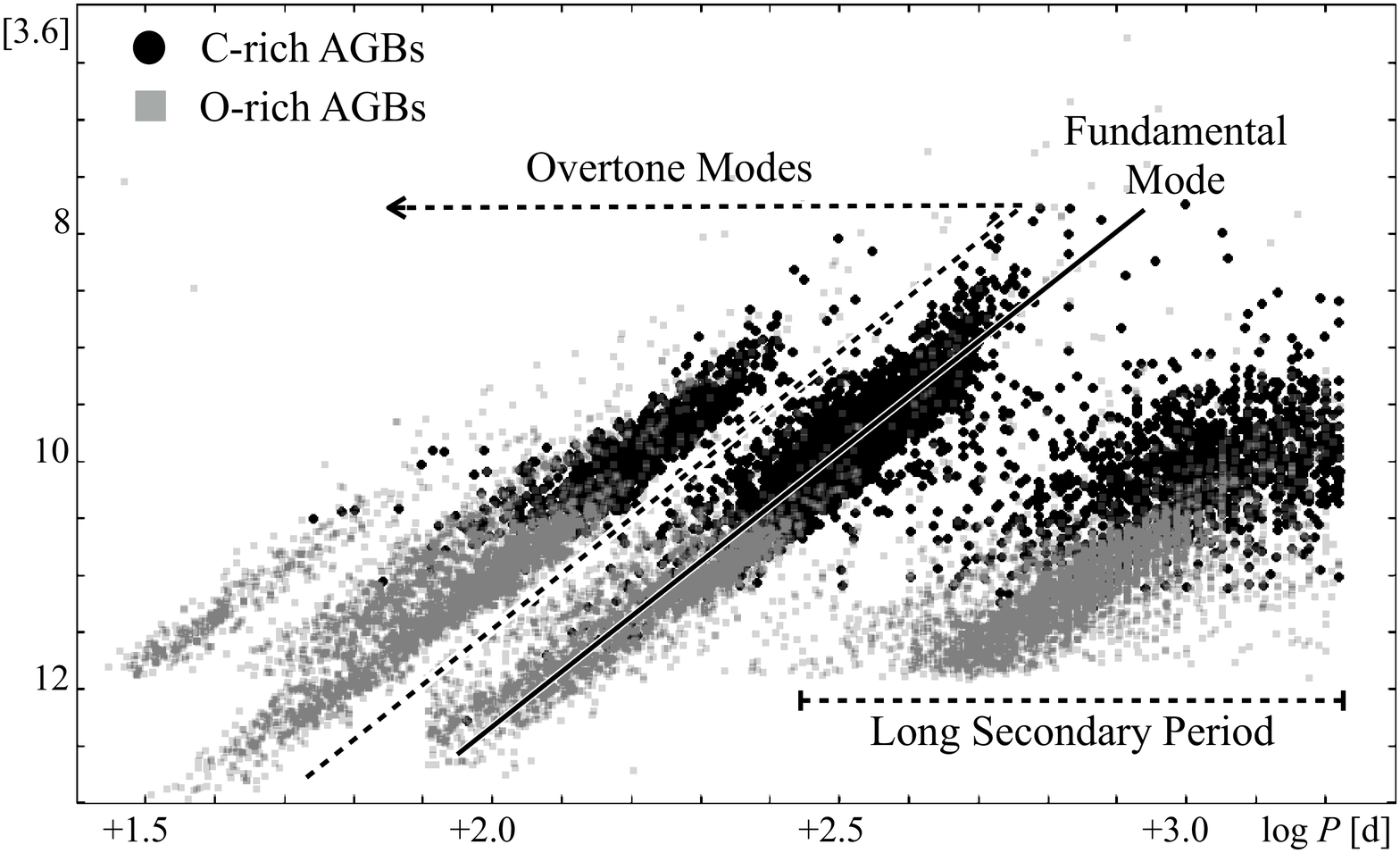}{fig:modes}
{Period-Luminosity diagram of AGB stars in the LMC found by  
  \citet{riebel2012} by matching the SAGE catalog with MACHO  
  lightcurves. The C/O dust chemistry of each source was determined by  
  radiative transfer fitting of its optical and infrared SED.}
  
The recent availability of deep photometric surveys covering entire
galaxies in the near-IR (e.g. the VISTA Magellanic Cloud survey, VMC;
\citealt{cioni2011}) and in the mid-IR (e.g. the ``Surveying the
Agents of a Galaxy's Evolution'', SAGE and SAGE-SMC surveys;
\citealt{meixner2006, gordon2011}) opened the possibility of deriving
dust mass loss rates $\dot M_d$ for entire populations of evolved
stars \citep{gullieuszik2012, riebel2012}. As an example,
Figure~\ref{fig:dmdt} shows the correlation between mid-IR colors and
$\dot M_d$ for AGB stars in the LMC, based on the model fitting of
SAGE data by \citet{riebel2012}. This approach, when applied
systematically to different environments within the Galaxy (e.g.
globular clusters and the bulge) and in galaxies within the local
group, provides a direct handle on the dependence of dust abundance
and mass loss rate from metallicity (see e.g. the ``DUST in Nearby
Galaxies with Spitzer'', DUSTiNGS, targeting 50 local group dwarf
galaxies with Spitzer; \citealt{boyer2014}). This technique is also
crucial to understand the yield of stellar mass loss to the overall
dust budget in galaxies \citep{matsuura2009, boyer2012, riebel2012,
  zhukovska2013}.

Optical surveys, meanwhile, have been targeting the time domain. The
MACHO \citep{alcock1996} and OGLE-III \citep{udalski2008} surveys, in
particular, have collected lightcurves for thousands of LPVs in the
Magellanic Clouds. Combined with infrared photometric surveys, these
lightcurves allow the characterization of evolved stars according to
their pulsation properties, establishing the link between variability
and mass loss. A typical example is shown in Figure~\ref{fig:modes},
where AGB stars in the LMC self-organize in separate Period-Luminosity
(PL) sequences based on their pulsation mode. These sequences, first
described by \citet{wood2010}, separate fundamental mode pulsators
from first and higher order overtone pulsators, as well as from a
class of ``secondary long period'' variables whose nature is
uncertain. The PL sequences in Figure~\ref{fig:modes} offer a more
quantitative diagnostics to characterize the pulsation properties of
LPVs than the traditional \emph{Mira} (which pulsate in fundamental
mode) vs. \emph{semiregular} (both fundamental and overtone)
dichotomy. The lack of reliable distances hinders similar analysis for
galactic LPVs (whose parallaxes are notoriously difficult to measure),
even though efforts have been made in this direction \citep[see
e.g.][]{tabur2010}.

The detailed physical processes responsible for mass loss, on the
other hand, can be uniquely studied using interferometric techniques
in the infrared (VLTI), sub-mm (SMA and now ALMA) and radio (EVLA,
VLBI). The high angular resolution achieved with interferometers
enables precision measurements in the circumstellar mo\-lecular layers
where the dust is condensed and the wind accelerated (see Wittkowski
review in this volume). These observations provide strong constrains
to models by spatially resolving the dust condensation sequence and
the asymmetries within the MOLsphere (the extended molecular gas
atmosphere) of pulsating giant stars \citep[see e.g.][]{karovicova2013,
  sacuto2013}, aided by the wealth of spectral data collected by
Herschel \citep[e.g.][]{lombaert2013, khouri2014}, and thermal infrared
spectrograph on ground-based telescopes and onboard SOFIA. Significant
progress is to be expected in this field in the near future, as
aperture synthesis image reconstruction in the near- and mid-IR is
coming to age, and as a new generation of VLTI instrument is being
developed (MATISSE, GRAVITY).

Finally, temporal variations in the mass loss rate are constrained by
wide field imaging of nearby stars, either in visible scattered light
\citep[e.g.][]{mauron2013}, in the UV \citep{martin2007, sahai2010},
in the infrared (most recently with Herschel, e.g. \citealt{cox2012}),
in the radio \citep{matthews2013} and in the sub-mm with ALMA
\citep{maercker2014}. These observations also provide unique
information on how the stellar outflow interacts with the ISM, and
allow resolving the degeneracy between morphology, temperature and
density in radiative transfer modeling, leading to more accurate
estimates of the mass loss rate.

\section{Mass Loss Modeling}
\label{sec:models}

A complete understanding of the physical principles behind mass loss
in evolved stars has proven to be elusive, due to the complex
interplay between molecular chemistry (some of which requires
non-LTE treatment, see e.g. \citealt{ryde2014}), dust condensation and
growth, non-linear dynamics of the pulsations and wavelength-dependent
radiative transfer.

Despite these difficulties, significant progress has been made in the
last decade. Pioneering dynamic models were based on a crude
treatment of grey radiative transfer \citep{bowen1988, hofner1997,
  winters2000}. Realistic spectra capable of reproducing the optical and
infrared observations of LPVs only appeared when models started to
include frequency-dependent radiative transfer \citep{hofner2003}, and
the complex chemistry network leading to dust condensation and growth
\citep[e.g.][]{gail1998}. While these models were initially using mass
loss as a boundary condition, in order to determine the emerging
spectra based on a prescribed dust production rate, the situation has
much improved in the latest generation of models. For the first time,
models are capable to produce synthetic mass loss rate predictions in
both C and O-rich environment \citep[e.g.][]{mattsson2008,
  mattsson2010, eriksson2014}, taking full advantage of the wealth of
observational constraints described in Section~\ref{sec:obs} and
allowing to explore the dependence of mass loss from stellar
parameters.

One point where dynamic models still lack fundamental physics is in
the treatment of pulsations, which are forced by introducing a
parametrized ``piston'' providing the boundary conditions at the base
of the atmosphere. While this treatment is very effective in
reproducing the light curves of individual LPVs, and allows exploring
the effects of variability on mass loss rate, the lack of physical
understanding of LPV pulsations is an obstacle for population
synthesis models where the pulsation parameters of the evolved stars
are not known a priori. Progress in this sense may come from 3D dynamic
models \citep[see][]{freytag2008, chiavassa2014}: while
these models don't yet include the physics of pulsations, they appear
to spontaneously generate radial shocks that propagate and levitate
the stellar atmosphere almost to the escape velocity. These results
offer some hope that once the proper physics is introduced in the
stellar layers that drive the pulsations, we may reach a better
understanding of the LPV instability strip, and its effects on evolved
stars mass loss.

\section{Mass Loss Parametrizations}
\label{sec:params}

Evolutionary models require a reliable mass loss prescription in order
to calculate the mass of the convective envelope on top of the
nuclear-burning shells. As mentioned in Section~\ref{sec:intro},
knowing the mass loss rate at any point during the RGB, AGB and RSG is
crucial to determine the ultimate fate of the star (WD or SN) and the
yield to the ISM.

In absence of dust enhancement the mass loss of evolved stars is
generally described by equating the wind kinetic energy with the
radiative flux and the gravitational energy. The original formulation
\citep{reimers1975} of such a law was derived by estimating the mass
loss rate and wind velocity from circumstellar lines
(Ca~\textsc{ii} H \& K) in a magnitude-limited sample of
optically-bright M giants: 

\begin{equation}
\dot M_R = 1.34 \times 10^{-5} \, \eta \, \frac{L^{3/2}}{M \,
  T^2_{eff}}
\end{equation}

\noindent
were $L$ and $M$ are expressed in solar units, $T_{eff}$ in K, $\dot M$ in
$M_\odot$/yr and the efficiency parameter $\eta$ is calibrated by
fitting the Horizontal Branch (HB) of globular clusters. As pointed
out by the author, \citeauthor{reimers1975}-type scaling relations are
not appropriate to describe the intense winds characteristic of
dust-enshrouded evolved stars in the TP-AGB phase. Modifications are
required to reach the high mass loss rates of dust enhanced winds
\citep[e.g.][]{blocker1995} and describe the increasing mass loss rate
of a star as it evolves along the AGB or supergiant branches.

\articlefigure[width=0.85\textwidth,angle=0]{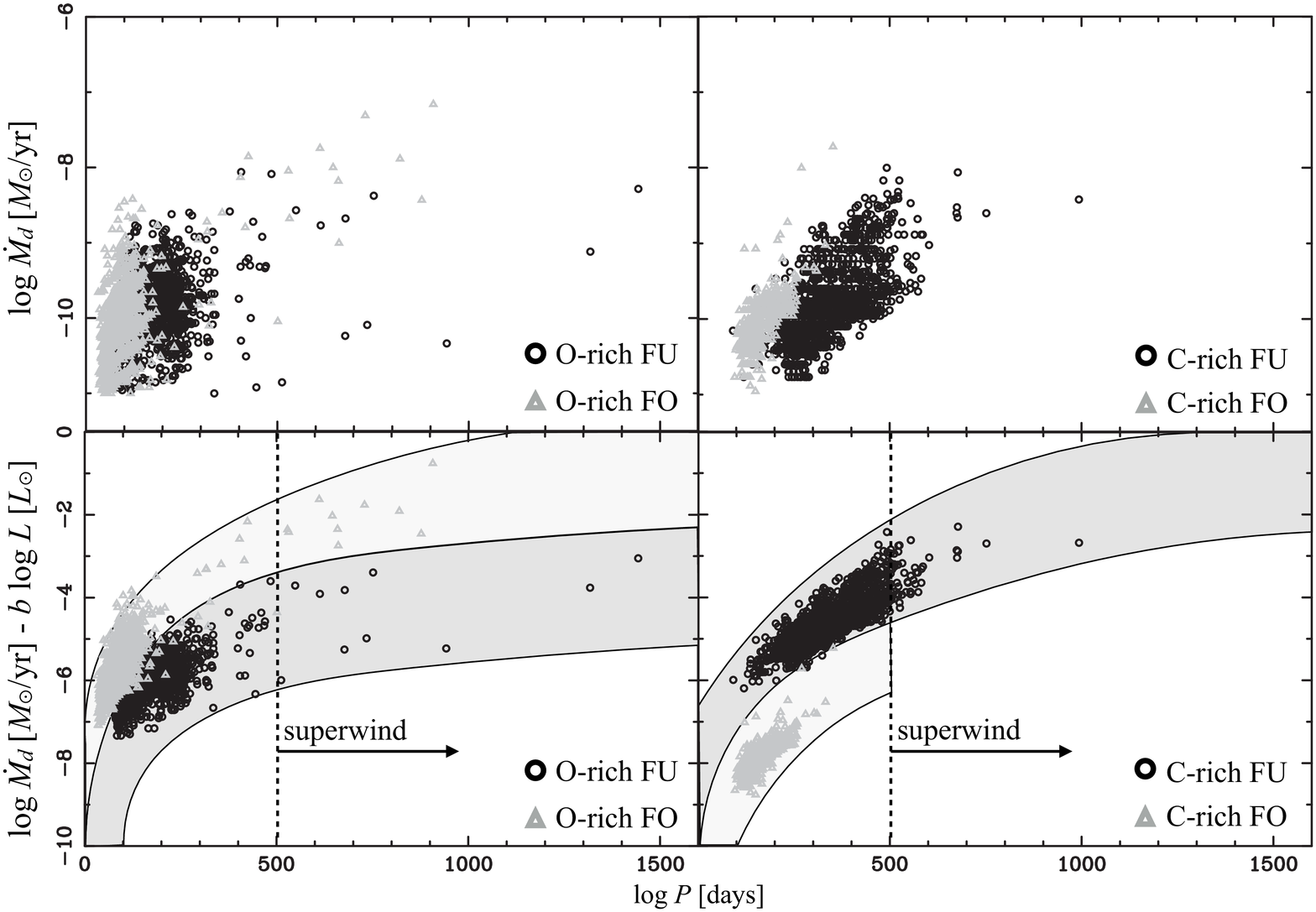}{fig:fit}
{Mass loss vs. period for LMC AGB stars, separately for O-rich
  (\emph{left}) and C-rich (\emph{right}) sources. Dark circles are
  variables selected from the fundamental mode sequence in
  Figure~\ref{fig:modes}, while grey triangles are selected from the
  first overtone sequence. The mass loss vs. period relation can be
  tightened by factorizing the luminosity along each pulsation
  sequence (\emph{bottom} panels).}

A different approach is followed in the \citet{vassiliadis1993} mass
loss prescription, that tackles the issue of mass loss during the
AGB phase. The mass loss rate was in this case determined from either
CO line emission (for LPVs with $P \la 500$~days) or from infrared
excess (IRAS data) for dust enshrouded variables (mostly OH/IR stars)
with longer periods. The break at $P \sim 500$~days is reflected in
the dependence of $\dot M$ from the pulsation period. For stars with
$P \la 500$~days $\dot M$ is a steep power law of the period

\begin{equation}
\left\{
\begin{array}{ll}
\log \dot M = -11.4 + 0.0123 \, P & \textrm{for} \ M \la 2.5 \ M_\odot\\
\log \dot M = -11.4 + 0.0125 \, [P - 100 (M/M_\odot - 2.5)] &
\textrm{for} \ M > 2.5 \ M_\odot 
\end{array} 
\right.
\end{equation}

\noindent
with $\dot M$ in $M_\odot$/yr and $P$ in days. For very long period
variables the $\log \dot M$ vs. $P$ relation flattens, with stars
entering a \emph{superwind} phase with $\dot M \simeq 10^{-4} \
M_\odot$/yr. In \citeauthor{vassiliadis1993} the dependence of the
mass loss rate from $R$ and $L$ is implicit, as derived from pulsation
theory \citep[e.g.][]{wood1990}.

Stellar population synthesis models use a variety of mass loss
prescriptions to parame\-trize mass loss in different evolutionary
phases. A ``state of the art'' example is found in
\citet{rosenfield2014}. A \emph{pre-dust} phase characteristic of the
RGB and early-AGB, when the wind is entirely pulsation-driven, are
described by modern formulations of the \citeauthor{reimers1975} wind,
modified to take into account the role of Alfv\'en waves, the residual
RGB envelope mass and MHD turbulence in convective zones
\citep{schroeder2005, cranmer2011}. Once dust appears the mass loss
switches to an \emph{exponential} mode, where the pulsation-driven
mass loss is enhanced by radiation pressure on the dust grains. This
is a \citeauthor{vassiliadis1993} wind of the form $\dot M \propto R^a
\, M^b$. The latest evolutionary phases on the AGB and RSG are then
described by a \emph{superwind} phase, as in \citet{vassiliadis1993},
characteristic of stars falling over the mass loss ``cliff'' at the
end of their life, rapidly depleting their convective envelope.

Much remains to be done. Current mass loss prescriptions do not
effectively take into account the different dust and molecular
chemistry in O-rich and C-rich evolved stars, despite evidence
that the third dredge-up does affect not just the infrared colors of
the circumstellar envelope, but also the mass loss rate \citep[see
e.g.][]{uttenthaler2013}. The large datasets provided by infrared
surveys such as SAGE, offer a unique possibility for determining a
robust parametrization of $\dot M$ according to the pulsation mode and
outflow chemistry. An example is shown in Figure~\ref{fig:fit},
showing $\log \dot M_d$ (from the \citealt{riebel2012} SED fitting)
vs. $P$ (\emph{top} panels). The relation can be significantly
tightened (\emph{bottom} panels) by factorizing the bolometric
luminosity of each source (again from \citealt{riebel2012} fit) with a
powerlaw $L^b$ (with the exponent $b$ determined individually for each
pulsation sequence and chemical type). Note that the artificial
break at $P \sim 500$~days, for superwind onset in the
\citeauthor{vassiliadis1993} mass loss, disappears in this parametrized
space, and the sources align on a single sequence $\log \dot M_d = a
\log P + b \log L$. Further work is however necessary to extend this
dataset (which depends on accurate period determinations from the
optical MACHO survey) to the most obscured sources (the extreme AGBs)
at long periods. Efforts to fill this gap by measuring the complete
lightcurve of very red LPVs in the LMC are the subject of a number of
current Spitzer programs (e.g. PID~10154, PI B. Sargent).

\section{Conclusions}
\label{sec:conclusions}

The availability of large scale photometric surveys has provided the
datasets required to explore the relation between mass loss  and
variability in complete samples of evolved stars in entire galaxies.
Combined with recent development in high resolution imaging and
spectroscopy of the molecular layers in mass losing stars, these
observational efforts are providing uniquely strong constraints for a
new generation of models capable to predict dust formation and mass
loss rates from the fundamental physics of cool pulsating atmospheres.
The ultimate goal is the generation of reliable mass loss
prescriptions for accurate stellar evolution and population synthesis
simulations.

\acknowledgements The author would like to thank Lee-Anne Willson for
the numerous conversations enjoyed while preparing this work and
Dieter Reimers for helping to improve this manuscript.

\bibliographystyle{asp2010}
\bibliography{marengo}

\end{document}